\documentclass[12pt]{article}
\usepackage{graphicx}
\usepackage[top = 2cm]{geometry}
\usepackage{hyperref}
\usepackage{xcolor}
\usepackage{amsmath}

\graphicspath{{figures/}}

\begin{document}

\def\onehalf{{\textstyle \frac12}}
\def\ii{{\rm i}}
\def\dd{{\rm d}}
\def\jour#1#2#3#4{{\it #1{}} {\bf #2}, #3 (#4)}
\def\lab#1{\label{eq:#1}}
\def\rf#1{(\ref{eq:#1})}
\def\Lie#1{\hbox{\sf #1}}
\def\FFF{{\sf F}}
\def\rectangulo#1{\centerline{\framebox{\LARGE #1}}}
\def\of#1{{\scriptstyle(}#1{\scriptstyle)}}
\def\oof#1{{\scriptscriptstyle(}#1{\scriptscriptstyle)}}
\def\tsty#1#2{{\textstyle\frac{#1}{#2}}}
\def\ssty#1{{{\scriptscriptstyle #1}}}
\def\ssr#1{{\scriptscriptstyle\rm #1}}
\def\abs#1{{\scriptstyle|}#1{\scriptstyle|}}
\def\vecdos#1#2{\bigg({#1 \atop #2}\bigg)}
\def\matdos#1#2#3#4{\bigg({#1 \atop #3}\ {#2\atop #4}\bigg)}
\def\mattres#1#2#3#4#5#6#7#8#9{\left(\begin{array}{ccc}
	#1&#2&#3\\ #4&#5&#6\\ #7&#8&#9 \end{array} \right)}	
\def\matricita#1#2#3#4{{\textstyle \Big({#1 \atop #3}\ {#2\atop #4}\Big)}}
\def\Re{{\rm Re}\,}  \def\Im{{\rm Im}\,}
\def\indket#1#2{| #1 \rangle_{\scriptscriptstyle #2}}
\def\indbra#1#2{{}_{\scriptscriptstyle #1}\langle #2 |}
\def\indbraket#1#2#3#4{{}_{\scriptscriptstyle #1}\langle #2 | #3 \rangle_{\scriptscriptstyle #4}}
\def\ket#1{|\,#1\,\rangle}
\def\roundket#1{|\,#1\,)}
\def\bra#1{\langle\,#1\,|}
\def\roundbra#1{(\,#1\,|}
\def\braket#1#2{\langle\,#1\,|\,#2\rangle}
\def\roundbraket#1#2{(\,#1\,|\,#2\rangle}
\def\braroundket#1#2{\langle\,#1\,|\,#2)}
\def\binom#1#2{\bigg({#1\atop #2}\bigg)}
\def\jour#1#2#3#4{{\it #1{}} {\bf #2}, #3 (#4)}
\def\lab#1{\label{eq:#1}}
\def\rf#1{(\ref{eq:#1})}
\def\Lie#1{\hbox{\sf #1}}
\def\rectangulo#1{\centerline{\framebox{\LARGE Figura #1}}}

\renewcommand{\thefootnote}{$\star$} 

\newcommand{\be}{\begin{equation}}
\newcommand{\ee}{\end{equation}}
\newcommand{\bea}{\begin{eqnarray}}
\newcommand{\eea}{\end{eqnarray}}


\begin{center}
{\LARGE Arbitrary unitary rotation of\\ three-dimensional pixellated images}\\[20pt] 	
 {Alejandro R.\ Urz\'ua} and {Kurt Bernardo Wolf\footnote{Deceased May 25, 2022}}\\[15pt]
{Instituto de Ciencias F\'{\i}sicas\\
Universidad Nacional Aut\'onoma de M\'exico\\Av.\ Universidad s/n, 
Cuernavaca, Morelos 62251, M\'exico}
\end{center} 

\vskip25pt

\begin{abstract}
Using the coefficients introduced by Bargmann and Moshinsky for the reduction
of the \Lie{su($3$)} algebra of Cartesian three-dimensional oscillator multiplet states
into \Lie{so($3$)} angular momentum submultiplets, we implement unitary rotations of 
three-dimensional Cartesian arrays that form finite pixellated ``volume images.''
Transforming between the Cartesian and spherical bases, the subgroup of 
rotations in the latter is converted into rotations of the former, 
allowing for proper concatenation and inversion of these unitary transformations,
which entail no loss of information.
\end{abstract}


\vskip1cm

\section{Introduction\label{sec:one}}  

Unitary rotations of three-dimensional ($3D$) pixellated `images' in a cube of $N^3$ volumetric data, will be \emph{imported} from the rotations of the $3D$ quantum harmonic oscillator wavefunctions between Cartesian and spherical coordinates onto the $3D$ \emph{discrete} oscillator model \cite{AW}. The importation process will follow closely the process which was applied in the $2D$ case of $N^{2}$ pixellated arrays or screens \cite{APVW-II,LEV-KBW}. This work relies heavily on the understanding of group theory, and although it is not a requirement, we do not rely on conventional notation in signal analysis; nevertheless, the reader can follow, without problem, the development of the presented methods

We start with the well-known $D$-dimensional Heisenberg-Weyl algebra of position and momentum operators $\overline{Q}_{i}, \overline{P}_{i}, \it{1} \in \Lie{hw}_{i}$, with their standard commutation relation $[\overline{Q}_{i}, \overline{P}_{j}]=\ii \delta_{i,j}\it{1}$. These serve to construct the raising and lowering operators,  $\overline{Q}_{i} \pm \ii\overline{P}_{i}$, out of which oscillator  Hamiltonians for each axis are built as the products $\overline{H}_{i} := \onehalf(\overline{P}{}^2_{i} + \overline{Q}{}^2_{i})$, also defining the \emph{number} operators $\overline{N}_{i} := \overline{H}_{i} - \onehalf\it{1}$ with integer eigenvalues $n_{i} \in \{0,1,2, \ldots\}$, as well as the quadratic product operators that shift quanta between the axes, which form the \emph{symmetry} Lie algebra \Lie{su($D$)} of the $D$-dim oscillator system. These operators conserve the total energy $E_{n}=\hbar\omega(n + \onehalf D)$, given by the principal quantum number $n = n_{1} + n_{2} + \cdots + n_{D}$. This \Lie{su($D$)} Lie algebra contains $D$-dimensional orthogonal subalgebras \Lie{so($D$)} of \emph{rotations}. 

The harmonic oscillator wavefunctions are separable in Cartesian coordinates and also in polar or, generally,  spherical coordinates. In the former, the quantum numbers $n_{i}$ are provided by the $D$ commuting number operators $\overline{N}_i$,  while the latter use quantum numbers from the rotation subalgebra chain. For $D=2$ dimensions, denoted $x, y$, the oscillator states are classified by
\begin{equation}
\begin{aligned}
	\overline{N}_{x}\oplus\overline{N}_{y}\quad \subset\quad \Lie{su($2$)} &= \Lie{so($3$)}\quad \supset\quad \Lie{so($2$)}\\
	n_{x}\hspace{2em} n_{y}\quad\quad n := n_{x} &+ n_{y} =:2\ell\qquad\quad m\\
	n_{i}\in\{0,1,\ldots\}, i\in\{x,y\}\qquad & \qquad m\in\{-\ell, -\ell+1,\ldots, \ell-1, \ell\}.
\end{aligned}
\end{equation} 
and form an inverted infinite triangle with apex at the ground state. At level $n = 2\ell$ there are $n + 1 = 2\ell + 1$ states degenerate  in energy $E_{n} = n + \onehalf$, that form an angular momentum multiplet of states classified by $m$. \cite{APVW-II,LEV-KBW}.

Correspondingly, for $D=3$ dimensions, we have the 
subalgebra chains and quantum numbers given by
\begin{equation}
\begin{aligned}
	\overline{N}_{x}\oplus\overline{N}_{y}\oplus\overline{N}_{z}\qquad\quad \subset\quad &\Lie{su($3$)}\qquad\quad \supset\Lie{so($3$)}\supset\Lie{so($2$)}\\
	n_{x}\hspace{2em} n_{y}\hspace{2em} n_{z}\quad\quad n := n_{x} &+ n_{y} + n_{z} \qquad\quad l\qquad m\\
	n_{i}\in\{0,1,\ldots\}, i\in\{x,y,z\}\qquad & \qquad \genfrac{}{}{0pt}{0}{l\in L(n) := \{n, n-2, \ldots, 0\;\mathrm{or}\; 1\}}{m\in M(\ell):= \{-\ell, -\ell+1,\ldots, \ell-1, \ell\}}.
\end{aligned}\lab{varranges}
\end{equation}
In this case, each energy level $n$ contains states that can be arranged in an inverted infinite triangular prism with apex at the  ground state. The multiplet of states with energy $E_{n} = n + 1$ forms triangular numbers $\tau_{n} := \sum_{0(+2)}^{n}(2\ell + 1) =  \onehalf(n + 1)(n + 2)$, which for $n\ge 2$ contain more than one angular momentum submultiplet of states. The linear combination coefficients that relate the Cartesian states $\ket{n_{x},n_{y},n_{z}}$ and the spherical $3D$ quantum oscillator states $\roundket{n, \ell, m}$ were given by Bargmann and Moshinsky in \cite{Barg-Mosh} for the nuclear shell model, and further investigated in \cite{Chacon-deLlano,Chasman}. We use a 
\emph{round} ket because $\ell$ is \textsc{not} the eigenvalue of any one operator in the set, but $\ell(\ell + 1)$, $\ell \ge 0$, is the eigenvalue of \emph{squared} total angular momentum.

The \emph{finite model oscillator} was proposed in \cite{AW,APVW-I} to form a system with oscillator dynamics but with a finite number $N$ of energy states. This was achieved from the Heisenberg-Weyl algebra by \emph{substituting} $\overline{Q}_{i}$ and $\overline{P}_{i}$ with the triads of operators $Q_{i}, P_{i}, K_{i}\in \Lie{su($2$)}$ with the commutation relations $[Q_{i}, P_{i}] = \ii K_{i}$, $[K_{i}, Q_{i}]=\ii P_{i}$, $[K_{i}, P_{i}] = -\ii Q_{i}$ and, in the irreducible representation $j\in\{0,\frac12,1,\frac32,\ldots\}$  of \Lie{su($2$)}, the `displaced finite oscillator number operators' $N_{i} := K_{i} + j{\it1}$ whose multiplets have $N := 2j + 1$ states with `number' and `energy' eigenvalues $n = E_{n}^\ssty{(N)} - \onehalf \in\{0,1,2,\ldots,N-1\}$.

Finally, the \emph{symmetry importation} from the continuous to the discrete model is performed by applying the rotation linear combination coefficients of the former, given by the Wigner Big-$D$ matrices $D^\ell_{m, m'}(\alpha, \beta, \gamma)$, onto the latter. 
 
In Sect.\ \ref{sec:two} we give a short account of the construction of the transformation coefficients for the continuous system; in Sect.\ \ref{sec:three} we import this symmetry onto the discrete cube of data, give the correct implementation of the explicit expression given by \cite{Chacon-deLlano}. In Sect.\ \ref{sec:four} we present some examples of unitary rotations and their concatenation.


\section{Transformation coefficients for\\ the $3D$ quantum oscillator}\label{sec:two}

To conform with the prevailing notations in the field, instead of the position and momentum operators, $Q_{i}$, $P_{i}$ we use the creation and annihilation operators\footnote{People used to the standard quantum notation can make the relations $\eta\equiv\hat{a}$ and $\xi\equiv\hat{a}^{\dagger}$.}
\begin{equation}
	\eta_i:= \frac1{\surd2}(Q_i+\ii P_i), \quad 
		\xi_i:= \frac1{\surd2}(Q_i-\ii P_i), \qquad
	[\xi_i,\eta_j]=\delta_{i,j}{\it1},\quad i,j\in\{x,y,z\}
			\lab{xi-eta-comrel}
\end{equation}
The three Cartesian commuting oscillator boson creation  operators $\vec\eta = (\eta_{x}, \eta_{y}, \eta_{z})$, and their three adjoint (hermitian conjugate) annihilation operators $\vec\xi = (\xi_{x}, \xi_{y}, \xi_{z})$, $\xi_{i} = \eta_{i}^{\dagger}$, act on the independent oscillator mode states, indicated by $\ket{n_{i}}$ in Dirac ket notation,
\begin{equation}
	\eta_i\ket{n_i}=\sqrt{n_i{+}1}\,\ket{n_i{+}1},\quad
	\xi_i\ket{n_i}=\sqrt{n_i}\,\ket{n_i{-}1},\qquad
		\eta_i\xi_i\ket{n_i}= n_i\ket{n_i}.
					\lab{rai-low-com}
\end{equation}
They serve to create the Cartesian oscillator states out of the ground state $\ket0$ as
\begin{equation}
	\ket{n_x,n_y,n_z}
			=\frac{\eta_x^{n_x}\eta_y^{n_y}\eta_z^{n_z}
							}{\sqrt{n_x!\,n_y!\,n_z!}}\ket0.
					\lab{Cart-sstates}
\end{equation}

Equivalently it is useful to define their three \emph{polar} complex linear combinations,
\begin{equation}
\begin{aligned}
		\eta_{+}: = -\frac1{\surd2}(\eta_{x} + \ii\eta_{y}),\quad \eta_{0} &:=\eta_{z},\quad \eta_{-} := +\frac1{\surd2}(\eta_{x} - \ii\eta_{y}),\\
		\xi_{+} := -\frac1{\surd2}(\xi_{x} - \ii\xi_{y}),\quad \xi_{0} &:=\xi_{z},\quad \xi_{-} := +\frac1{\surd2}(\xi_{x} + \ii\xi_{y}),
\end{aligned} 					\lab{eta-xi-polar}
\end{equation}
where $\xi_{r} = \eta_{r}^{\dagger}$ and $[\xi_{r},\eta_{s}] = \delta_{r,s}{\it1}$ and $r, s\in\{+,0,-\}$. The inverse transformation is
\begin{equation}
\begin{aligned}
		\eta_{x} := -\frac1{\surd2}(\eta_{+} - \eta_{-}),\quad \eta_{y} &:= +\ii\frac1{\surd2}(\eta_{+} + \eta_{-}),\quad \eta_{z} := \eta_{0},\\
		\xi_{x} := -\frac1{\surd2}(\xi_{+} - \xi_{-}),\quad \xi_{y} &:= -\ii\frac1{\surd2}(\xi_{+} + \xi_{-}),\quad \xi_{z} := \xi_{0},		
\end{aligned}	\lab{eta-xi-cart}
\end{equation}
The polar operators act, as the Cartesian ones in \rf{rai-low-com}, on  states $\ket{n_r}^{\!\circ}$, $r\in\{+,0,-\}$, through
\begin{equation}
	\eta_r\ket{n_r}^{\!\circ}=\sqrt{n_r{+}1}\,\ket{n_r{+}1}^{\!\circ},\ 
	\xi_r\ket{n_r}^{\!\circ}=\sqrt{n_r}\,\ket{n_r{-}1}^{\!\circ},\ \ 
		\eta_r\xi_r\ket{n_r}^{\!\circ}= n_r\ket{n_r}^{\!\circ}.
					\lab{rai-low-com}
\end{equation}
Thus one defines the \emph{polar mode} states
\begin{equation}
			\ket{n_+,\,n_0,\,n_-}^{\!\circ} 
				:= \frac{\eta_+^{n_+}\eta_0^{n_0}\eta_-^{n_-}
							}{\sqrt{n_+!\,n_0!\,n_-!}}{\ket0}^{\!\circ} ,
							\lab{polarkets}
\end{equation}
with the unique ground state $\ket0^{\!\circ}\equiv{\ket0}$.

The Cartesian and polar states thus relate through
\begin{equation}
\begin{aligned}
	\ket{n_x,n_y,n_z}
			&=\frac{\ii^{n_y}(\eta_-{-}\eta_+)^{n_x}(\eta_-{+}\eta_+)^{n_y}\eta_0^{n_z}
							}{\sqrt{2^{n_x+n_y}\,n_x!\,n_y!\,n_z!}}\ket{0}\\
				&= \ii^{n_y}\sqrt{\frac{n_x!\,n_y!}{2^{n_x+n_y}}}
						\sum_{\lambda=0}^{n_x}\sum_{\mu=0}^{n_y} (-1)^\lambda\\
						&{}\times\frac{\sqrt{(\lambda{+}\mu)!\, (n_x{+}n_y{-}\lambda{-}\mu)!}
									}{\lambda!\,\mu!\,(n_x{-}\lambda)!\,(n_y{-}\mu)!} 
							{\ }\ket{\lambda{+}\mu,\,n_z,\,n_x{+}n_y{-}\lambda{-}\mu}^{\!\circ},
\end{aligned} \lab{Cartketexpr0}
\end{equation}
and the inverse relation
\begin{equation}
\begin{aligned}
	\ket{n_+,n_0,n_-}^{\!\circ} 
			&=\frac{(-1)^{n_+}(\eta_x{+}\ii\eta_y)^{n_+}\eta_z^{n_0}(\eta_x{-}\ii\eta_y)^{n_-}
							}{\sqrt{2^{n_++n_-}\,n_+!\,n_0!\,n_-!}}\ket0^{\!\circ}\\
				&= (-\ii)^{n_++n_-}\sqrt{\frac{n_+!\,n_-!}{2^{n_++n_-}}}
						\sum_{\rho=0}^{n_+}\sum_{\sigma=0}^{n_-} \ii^{\sigma-\rho} \\
					&{}\times	\frac{\sqrt{(n_+{+}n_-)!\, (n_+{+}n_-{-}\rho{-}\sigma)!}
									}{\rho!\,\sigma!\,(n_+{-}\rho)!\,(n_-{-}\sigma)!} 
				{\ } \ket{n_+{+}n_-,\,n_+{+}n_-{-}\rho{-}\sigma,\,n_0}.
\end{aligned}\lab{Polketexpr0}
\end{equation}

The corresponding Schr\"odinger wavefunction $\braket{\vec r}{n_{x}, n_{y}, n_{z}}$, $\vec r = (x,y,z)$, are the well-known separated Hermite-Gauss functions, a denumerable and orthonormal basis for functions in the ${\cal L}^2({\sf R}^3)$ Hilbert space. 

In the spherical basis, the states are determined by \cite{Chacon-deLlano,Davies}, \cite[Sects.\ I.8-9]{Moshinsky-Smirnov} 
\begin{equation}
		\roundket{n,\ell,m} := A_{n,\ell}\,
				(\eta^2)^{\frac12(n-\ell)}\,
							{\cal Y}_{\ell,m}(\vec{\,\eta})\ket0,
\end{equation}\lab{sph-basis}
where $\eta^{2} := \eta^{2}_x{+}\eta^{2}_y{+}\eta^{2}_z = \eta_{0}^{2} - 2\eta_{+}\eta_{-}$ is raised to the power $\nu := \onehalf(n - \ell)$; $\nu$ is a non-negative integer identified as the \emph{radial quantum number}. The ${\cal Y}_{\ell,m}(\vec{\,\eta})$'s are the \emph{solid spherical harmonics} in the three creation operators $\vec\eta$;  written in polar form \rf{eta-xi-polar}, these are \cite[Sect.\ 3.10]{Bied-Louck},
\begin{equation}
\begin{aligned}
	{\cal Y}_{\ell,m}(\vec{\,\eta}) 
		&= \displaystyle \sqrt{\frac{(2\ell{+}1)\,(\ell{+}m)!\,(\ell{-}m)!}{4\pi\, 2^m}}\\
			& \displaystyle \qquad{}\times\sum_{\mu} \frac1{2^\mu}\,
						\frac{\eta_+^{m+\mu}}{(m+\mu)!}\,
												\frac{\eta_0^{\ell-m-2\mu}}{(\ell-m-2\mu)!}\,
												\frac{\eta_-^{\mu}}{\mu!}\,.
\end{aligned}\lab{Solid-sph}
\end{equation}
When $m\ge 0$, then $\mu$ ranges from 0 to $\lfloor\frac12(\ell-m)\rfloor$; when $m<0$, we use ${\cal Y}_{\ell,-m}(\vec{\,\eta})=(-1)^m {\cal Y}_{\ell,m}(\vec{\,\eta})^*$. This is a polynomial of homogeneous degree $\ell$ in the creation operators. Together with the factor   $(\eta^2)^\nu$ in \rf{sph-basis}, we have a polynomial operator of degree $n=2\nu + \ell$ that creates oscillator states of energy  $E_{n}=\hbar\omega(2\nu{+}\ell{+}\frac32)$. Finally, the normalization coefficient in \rf{sph-basis} is \cite[Eqs.\;(11)]{Barg-Mosh}\footnote{Please note that Chac\'on and de Llano use ``$\nu$'' for the total quantum number $n$, while we use it for the radial quantum number in expressions where it simplifies notation.}
\begin{equation}
		A_{n,\ell}:=(-1)^{\frac12(n-\ell)}
				\sqrt{\frac{4\pi}{(n-\ell)!!\,(n{+}\ell{+}1)!!}}
				=(-1)^{\nu}
				\sqrt{\frac{4\pi}{(2\nu)!!\,(2\nu{+}2\ell{+}1)!!}}.
\end{equation}\lab{norm-coef}
We use the principal quantum number $n$, rather than the radial $\nu$, for our notation of bra-kets and coefficients  (in contradistinction to \cite{Barg-Mosh} and \cite{Pei-Liu-1}) because it simplifies many analytical and numerical consideratios. With $n=2\nu+\ell$, the spherical kets \rf{sph-basis}--\rf{norm-coef}  can be written in terms of the polar ones \rf{polarkets} as \cite{Chacon-deLlano}
\begin{equation}
\begin{aligned}
	\roundket{n,\ell,m} &= \frac{(-1)^\nu}{2^\ell\,\ell!}
		\sqrt{\frac{(2\ell{+}1)\,2^m\,(\ell{-}m)!}{(n{-}\ell)!!\,(n{+}\ell{+}1)!!}}\\
				&\phantom{\frac{(-1)^\nu}{2^\ell\,\ell!}}
					{}\times \sum_{s=0}^\ell \sum_{r=0}^{\nu+s}
				(-1)^{r+s}2^r\bigg({\ell\atop s}\bigg)\bigg({\nu{+}s \atop r}\bigg)
					\frac{(2\ell{-}2s)!}{(\ell{-}2s{-}m)}\\
				&\phantom{\frac{(-1)^\nu}{2^\ell\,\ell!}
					\sum_{s=0} \sum_{r=0}
				(-1)^{r+s}2^r}\times \ket{r+m,\,n-2r-m,\,r}^{\!\circ},
\end{aligned}\lab{roundexpr}
\end{equation}
for $m\ge0$, while for $m<0$, the expression for $\roundket{n,\ell,-m}$ in \rf{roundexpr} will exchange $\ket{n_+,\,n_0,\,n_-}^{\!\circ}\mapsto\ket{n_-,\,n_0,\,n_+}^{\!\circ}$. The Schr\"odinger wavefunctions $\braroundket{\vec r}{n,\ell,m}$,  are well known to be separated into a power and associated  Laguerre-Gauss function of the radius, times a spherical  harmonic of the direction of $\vec r$; these form also a  denumerable and orthonormal basis for ${\cal L}^2({\sf R}^3)$. We use round kets for the spherical basis to distinguish them when labeled with numerical values. 

In \cite{Chacon-deLlano} the overlap between the Cartesian and spherical bases, \rf{Cartketexpr0} and \rf{sph-basis} is reported. It is nonzero only when they belong to the same energy level and have the same parity in the $x$--$y$ plane, i.e.,
\begin{equation}
	n_x+n_y+n_z=n,\quad \hbox{and}\quad n_x+n_y\pm m \hbox{ is even}.
\end{equation}\lab{conditions}
Their expression, with $\nu=\onehalf(n-\ell)$, can be written as
\begin{equation}
\begin{aligned}
	\braroundket{n_x,n_y,n_z}{n,\ell,m}
			&=\ii^{n_y}(-1)^{\nu+\frac12(n_x-n_y\pm m)} 
						2^{-\ell}\,(\onehalf[n_x{+}n_y{+}m])!\\
			&{}\times\sqrt{(2\ell{+}1)\frac{(\ell{-}m)!}{(\ell{+}m)!}
						\frac{n_x!\,n_y!\,n_z!}{(n{-}\ell)!!\,(n{+}\ell{+}1)!!}}\\
			&\times\sum_{r=0}^\ell\sum_{s=0}^{n_x}
				\frac{(-1)^{r+s}\, (\nu+r)!\,(2\ell-2r)!
						}{{s!\,(n_x{-}s)!\,(\frac12[n_x{+}n_y{+}m]-s)!\,(\frac12[n_y{-}n_x{-}m]+s)! 
						\atop r!\,(\ell{-}r)!\,(\ell{-}2r{-}m)!\,(\frac12[n_z{-}\ell{+}m]+r)!}},
\end{aligned}\lab{Elbracket}
\end{equation}
for $m\ge0$, while for $m<0$,
\begin{equation}
		\braroundket{n_x,n_y,n_z}{n,\ell,m} 
			= (-1)^{n_x}\braroundket{n_x,n_y,n_z}{n,\ell,-m}.
\end{equation}\lab{Elbraketconm}
Note that all factorials in \rf{Elbracket} are applied to integer numbers due to \rf{conditions}.

The overlap coefficients \rf{Elbracket} transform unitarily between the $\tau_n$ Cartesian states within each energy level $n$, which form a basis for a completely symmetric irreducible representation of \Lie{SU($3$)}, and the spherical states which reduce into \Lie{SO($3$)}  angular momentum multiplets, and which sum the same number of states, $\sum_{\ell\in L(n)}(2\ell{+}1)=\tau_n$. For each $n\in\{0,1,2,\ldots\}$, the transformation relations between bases are
\begin{equation}
\begin{aligned}
	\roundket{n,\ell,m} &= \!\!\!\sum_{n_x+n_y+n_z=n}\!\!\!
			\ket{n_x,n_y,n_z}\braroundket{n_x,n_y,n_z}{n,\ell,m},\\				
	\ket{n_x,n_y,n_z} &= \!\!\sum_{\ell\in L(n),\ m\in M(\ell)}\!\!
			\roundket{n,\ell,m}\roundbraket{n,\ell,m}{n_x,n_y,n_z}.		
\end{aligned}\lab{xyz-sph}
\end{equation}
Now, in the spherical basis, $3D$ \emph{rotations}  $R\in\Lie{SO(3)}$ are transformations that only act \emph{within} the angular momentum multiplets 
\begin{equation}
\begin{aligned}
		  R:\roundket{n,\ell,m}&=\displaystyle
					  \sum_{m'\in M(\ell)}
			\roundket{n,\ell,m'}\roundbra{n,\ell,m'}\,
					R\,\roundket{n,\ell,m} \\[10pt] 
			&=\displaystyle\sum_{m'\in M(\ell)}
			      \roundket{n,\ell,m'}\,D_{m',m}^\ell(R),
\end{aligned}\lab{SO3trasf}
\end{equation}
where $D_{m',m}^\ell(R)$ are the well known Wigner \emph{big-D} rotation matrices, usually expressed in Euler angles $R(\phi,\theta,\psi)$.


\section{Importation of \Lie{SU($3$)} symmetry on the\\ finite $3D$ oscillator model}\label{sec:three}
					
The gist of the $1D$ finite oscillator model is to replace the Heisenberg-Weyl (\Lie{HW}) Lie algebra of raising ($\eta$) and lowering ($\xi{\,\equiv\,}\eta^\dagger$) operators
\begin{equation}
	\{\eta,\,\xi,\,{\it1}\} \in \hbox{span \Lie{HW}},
	         \qquad [\xi,\eta]={\it1},   
\end{equation}\lab{HW-algebra}
with an \Lie{su($2$)} algebra of generators $\{J_k\}_{k=1}^3$ in a fixed representation $j$, of dimension $N=2j+1$, and classify the basis states with the unit-spaced eigenvalues of either position or mode, as 
\begin{equation}
\begin{aligned}
	\hbox{position $q$:}\quad J_{1}|j,q\rangle_{1} &= q\,|j,q\rangle_{1}, q|_{-j}^j,\\
	\hbox{mode $n$:}\quad J_3|j,n\rangle_{3} &= (n{-}j)\,|j,n\rangle_{3}, n|_0^{N-1}.
\end{aligned}\lab{spectra}
\end{equation}
There are thus only $N$ mode eigenstates $n\in\{0,1,\ldots,N{-}1\}$  in the $1D$ finite oscillator model \cite{AW}. The \Lie{su($2$)} commutation relations are the usual ones \cite{Bied-Louck},
\begin{equation}
	[J_3,J_1]=\ii J_2,\quad [J_3,-J_2]=\ii J_1, \quad [J_1,J_2]=\ii J_3.
\end{equation}\lab{three-J-comm}
The first two are the geometric and dynamic Hamilton equations of the quantum oscillator, while the third distinguishes between the finite oscillator from the continuous model. The overlaps between the position and mode bases constitute the finite oscillator eigenstates found accross the related literature \cite{AW, Urz2016}, 
\begin{equation}
\begin{aligned}
	{\!\!\!}\Psi_n^{(j)}(q) &:= {}_1\langle j,q|j,n\rangle_3 
			= d^j_{n-j,q}(\onehalf\pi)\\
		&= \frac{(-1)^n}{2^j} \sqrt{\binom{2j}{n}
			\binom{2j}{j{+}q}}\,K_n(j{+}q;\onehalf;2j),
\end{aligned}
\end{equation}\lab{Krav-expr}
where $d^j_{m,m'}(\onehalf\pi)$ are the Wigner \emph{little-d} functions \cite{Bied-Louck} for the $\onehalf\pi$ angle between $J_1$ and $J_3$, and $K_n(j{+}q;\onehalf;2j)$ are the symmetric Kravchuk polynomials of degree $n$ in $q|_{-j}^j$. When $j\to\infty$ in an appropriate limit, the Kravchuk functions \rf{Krav-expr} become the Hermite-Gauss eigenfunctions of the continuous quantum oscillator.

\begin{figure}[htbp]
\centerline{\includegraphics[width=0.9\columnwidth]{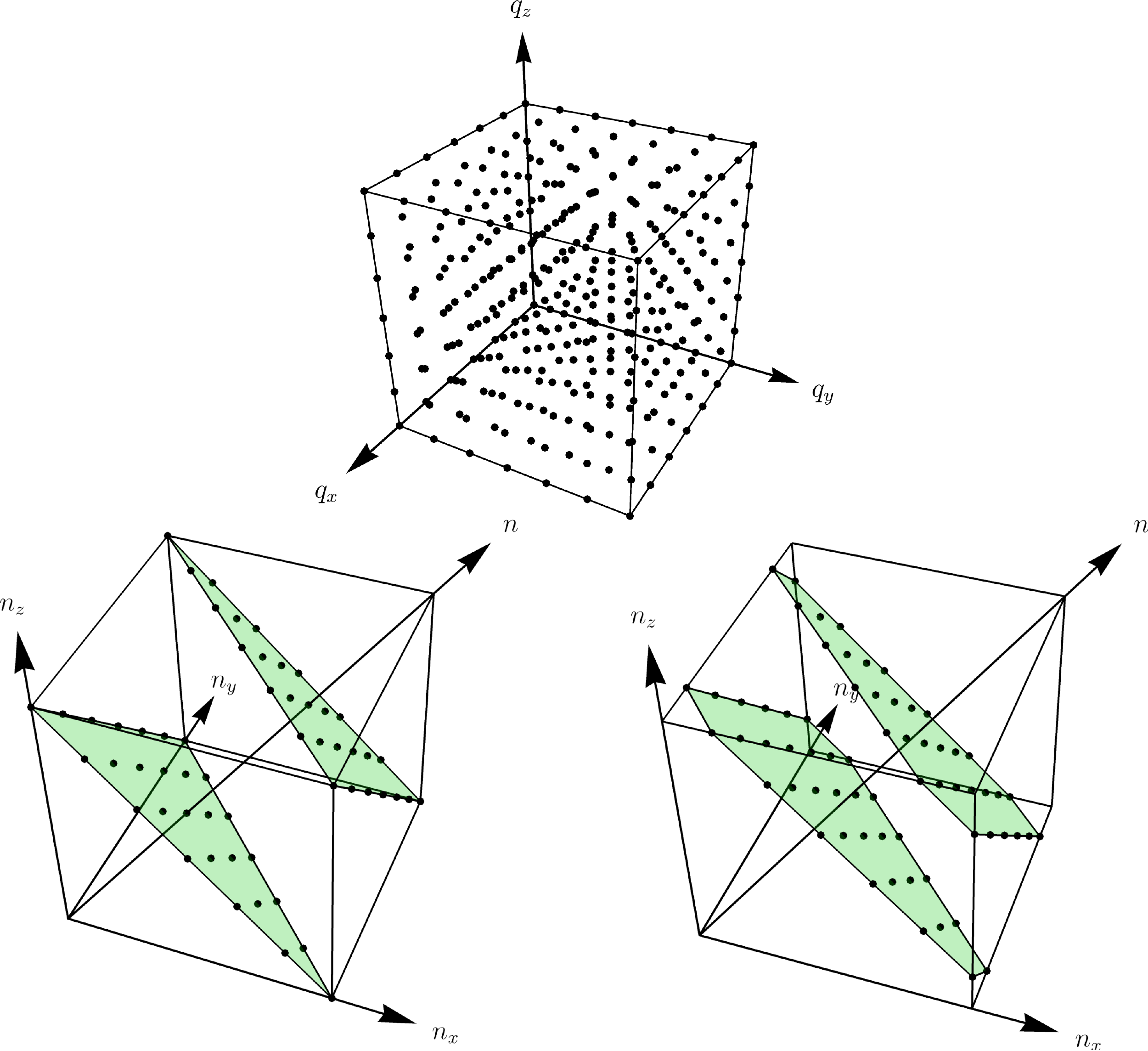}}
\caption[]{\textit{Above}: States of the $3D$ Cartesian finite oscillator wavefunctions $\Psi_{n_x,n_y,n_z}^\ssty{(N)}(q_x,q_y,q_z)$ in \rf{Psi-3D}, of points $q_x,q_y,q_z|_{-j}^j$, arranged into a cube of side $N=2j+1$. \textit{Below}: Eigenstates of mode number, arranged by axes $(n_x,n_y,n_z)$, with mode numbers $n_i|_0^{N-1}$. The diagonal axis crossing the cube distinguishes the total modes $n=\sum_i n_i$, the  right-down vertex is $n=0$ and the left-top vertex corresponds to $n=3(N{-}1)=6j$. \textit{Below-left}. Triangular planes are drawn between the lower (inverted) pyramid $n|_0^2j$, and the upper pyramid $n|_{4j}^{6j}$. \textit{Below-right}. Hexagonal planes are drawn for the intermediate region $n|_{2j+1}^{4j-1}$.} 
\label{fig:dos-cubos}
\end{figure}

In the three dimensions that we now study, we have three sets of commuting $\Lie{su($2$)}_i$ generators, $i\in\{x,y,z\}$, each giving position and mode \rf{spectra} along the three space axes. We are here interested in the case where $j_{x} = j_{y} = j_{z} =: j$ so that we have a total of $N^{3} = (2j + 1)^{3}$ states that we can represent as points in a cube of side $N$, which we picture in Fig.\ \ref{fig:dos-cubos} (above). Modifying slightly the notation in \rf{Krav-expr} where subscripts denotes the related basis, we write the discrete and finite position states as $\ket{q_{x}, q_{y}, q_{z}}_1^\ssty{(N)}$. On the other hand, the discrete mode states, correspondingly written as $\ket{n_{x}, n_{y}, n_{z}}_3^\ssty{(N)}$ with $n_{i}|_{0}^{N-1}$, are pictured in Fig.\ \ref{fig:dos-cubos} (below left-right) with the total mode number $n = n_{x} + n_{y} + n_{z}$ along the vertical axis, which ranges from $n=0$ at the bottom vertex, up to $n=3(N - 1) = 6j$ at the top vertex. A basis of $N^3$ discrete Cartesian wavefunctions is built from the $1D$ model in \rf{Krav-expr} simply as the direct product of the individual one-dimensional wavefunctions
\begin{equation}
\begin{aligned}
    \Psi_{n_x,n_y,n_z}^{(j)}(q_x,q_y,q_z) 
				&:= {}^\ssty{(N)}_{\phantom{N}1}\braket{q_x,q_y,q_z}{n_x,n_y,n_z}_3^\ssty{(N)}\\
				&\phantom{:}=\Psi_{n_x}^{(j)}(q_x)\Psi_{n_y}^{(j)}(q_y)\Psi_{n_z}^{(j)}(q_z).
\end{aligned}\lab{Psi-3D}
\end{equation}
These are orthonormal and complete bases for the $N^{3}$ space of `$3D$ images', `voxels', `signals' or \emph{states} on the discrete cube of Fig.\ \ref{fig:dos-cubos}. These act as unitary transformation matrix elements between the two bases depicted in that figure, i.e., 
\begin{equation}
\begin{aligned}
	 \ket{n_x,n_y,n_z}_3^\ssty{(N)} &=\!\!\!\!\!\! \sum_{q_x,q_y,q_z=-j}^j\!\!\!\!\!
						\ket{q_x,q_y,q_z}^\ssty{(N)}_{1}\, 
				{}^\ssty{(N)}_{\phantom{N}1}\!\braket{q_x,q_y,q_z}{n_x,n_y,n_z}_3^\ssty{(N)},\\
		\ket{q_x,q_y,q_z}_1^\ssty{(N)} &= \!\!\!\!\!\!\sum_{n_x,n_y,n_z=0}^{2j}\!\!\!\!\!
						\ket{n_x,n_y,n_z}^\ssty{(N)}_{3}\, 
				{}^\ssty{(N)}_{\phantom{N}3}\!\braket{n_x,n_y,n_z}{q_x,q_y,q_z}_1^\ssty{(N)}.
\end{aligned}\lab{ketq-ketn}
\end{equation}
where we remark that the basis for the modes $n_{i}$ is not so intuitive for visual representation, despite it has the same exact information that the basis of position $q_{i}$.

\begin{figure}[t]
\centerline{\includegraphics[width=0.9\columnwidth]{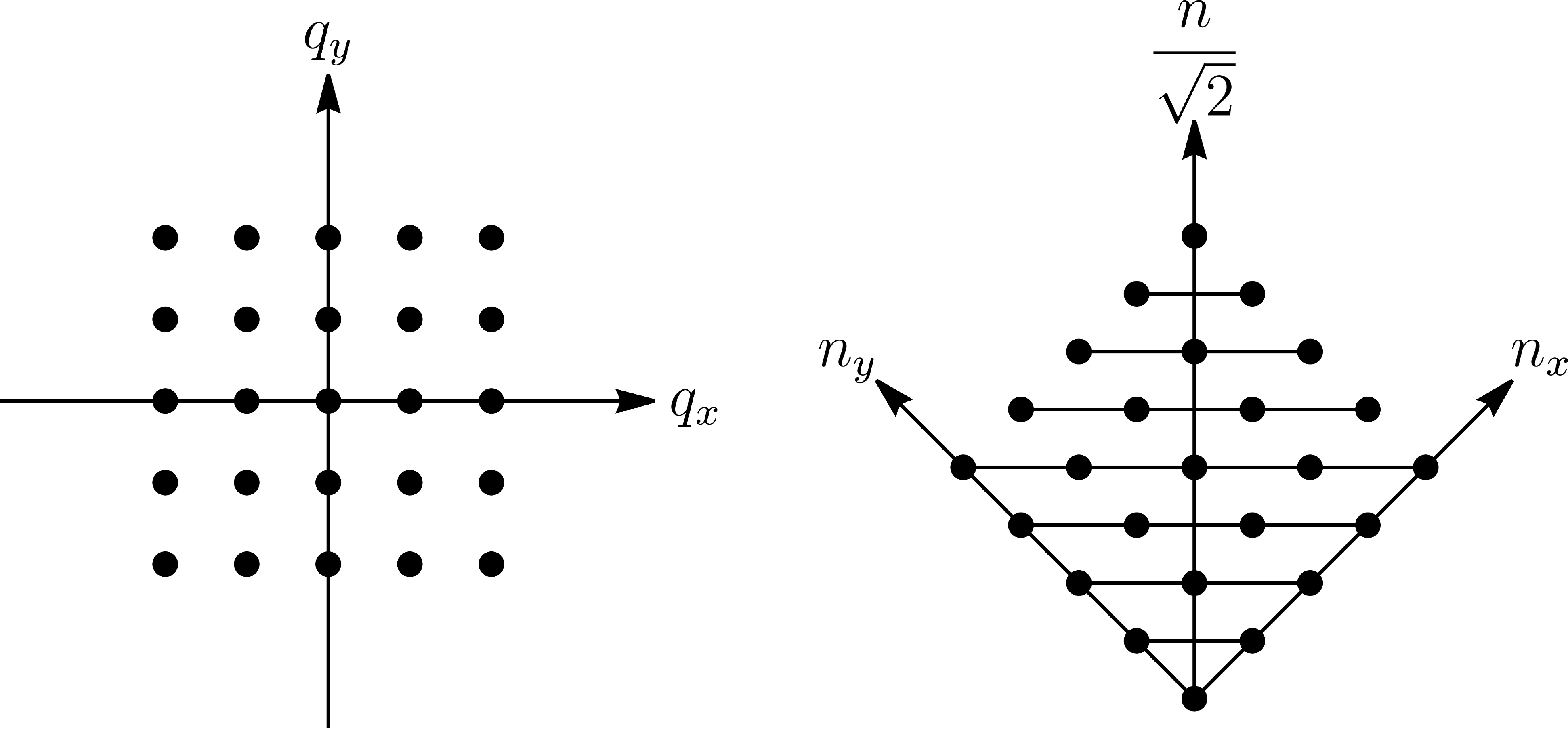}}
\caption[]{{\it Left\/}: States of a two-dimensional finite
oscillator represented as points (or pixels) on a square 
$N\times N$ screen ($N=2j{+}1$). \quad
{\it Right\/}: Mode states of the $2D$ finite oscillator, 
$\ket{n_x,n_y}_3$, $n_x,n_y|_0^{2j}$, and $n_x{+}n_y=n|_0^{4j}$ 
arranged into a rhombus. The horizontal lines join the 
{\it imported\/} \Lie{su($2$)} multiplets 
$\roundket{\lambda,\mu}$ of angular momenta 
$\lambda=\onehalf n\in\{0,\onehalf,1,\ldots,j\}$ in the
lower (and correspondingly in the upper) half of the rhombus, 
distinguished by $\mu=\onehalf(n_x-n_y)|_{-\lambda}^\lambda$.}
\label{fig:caso2D}
\end{figure}

In the \emph{two}-dimensional case examined in Refs.\ \cite{APVW-II,LEV-KBW}, the position states are placed on an $N\times N$ square pixellated screen as shown in Fig.\ \ref{fig:caso2D} (instead of the cube in Fig.\ \ref{fig:dos-cubos}), 
and the mode states are arranged in an $N\times N$ rhombus whose vertical axis is the total mode number $n=n_{x} + n_{y}$ and the horizontal axis is the mode difference $n_{x} - n_{y}$. The horizontal rungs in the rhombus, characterized by $n|_{0}^{2N - 2}$, contain each $n+1$ members in the triangle that is the lower half of the rhombus $0\le n\le N{-}1=2j$, while the upper triangle $N\le n\le2N{-}2=4j$ contains the high-mode components, whose members reflect the lower triangle. At this point, the \emph{importation} of the \Lie{su($2$)} symmetry consists in taking these sets of $n+1$ states to be \Lie{su($2$)} \emph{multiplets}, $\roundket{\lambda,\mu}$, of angular momentum $\lambda=\onehalf n$ and distinguished by the mode \emph{difference} $\mu=\onehalf(n_{x} - n_{y})$, $\mu|_{-\lambda}^\lambda$. This holds for the states in the triangle in the lower half of the rhombus; the upper triangle is treated in the same way for $n\mapsto 4j-n$. Thereby, plane rotations by $\alpha$ in the $x$--$y$ plane of the pixellated screen are obtained from the well-defined rotations of the $\roundket{\lambda,\mu}$ basis states, which are only multiplied by $\exp(\ii\alpha\mu)$ . 

Returning to the \emph{three}-dimensional case examined here, we must import the continuous oscillator \Lie{su($3$)} symmetry algebra onto the finite oscillator mode states at each total mode level $n = n_{x} + n_{y} + n_{z}$ of Fig.\ \ref{fig:dos-cubos}; and then use its \Lie{so($3$)} subalgebra to define states which transform properly under rotations  as in Eq.\ \rf{SO3trasf}. In this case we have to slice the rhomboid-cube of modes into $n={}$ constant planes and see if and how we can accommodate them into $\Lie{so($3$)}\subset\Lie{su($3$)}$ multiplets. In slicing the cube  in this way, we have now \emph{three} regions to examine, instead of the two triangles of the $2D$ rhombus in Fig.\ \ref{fig:caso2D}. 

Consider first the mode levels in the range $0\le n\le N-1$ where the slices of the cube yield triangular arrangements of $\tau_{n}$ points of side $n + 1$, such as that shown in Fig.\ \ref{fig:dos-cubos}. These form \emph{completely symmetric} (or bosonic) irreducible representations of \Lie{su($3$)}, containing the multiplet of the states  $\ket{n_{x}, n_{y}, n_{z}}$ of an ordinary quantum oscillator with $n$ energy quanta. These quanta can be shifted among the three axes by means of the nine \Lie{u($3$)} generators. These generators, ${\cal C}_{i}^{j}$, $i, j\in\{x, y, z\}$, are built in terms of three commuting sets of boson creation and annihilation operators, $\eta_{i},\,\xi_{j}$, fulfilling the commutators
\begin{equation}
		{\cal C}_i^j := \eta_i\xi_j,\qquad [{\cal C}_i^j,{\cal C}_k^l]
           =\delta_{j,k}\,{\cal C}_i^l-\delta_{l,i}\,{\cal C}_k^j,
					\qquad ({\cal C}_i^j)^\dagger= {\cal C}_j^i.
\end{equation}\lab{su3gens}
In the vector space of these nine generators there is the invariant of total mode 
\begin{equation}
	{\cal C}:=\sum_i {\cal C}^i_i=\eta_x\xi_x +\eta_y\xi_y +\eta_z\xi_z 
				= \eta_+\xi_+ + \eta_0\xi_0  + \eta_-\xi_-,  
\end{equation}\lab{Ccenter}
which is at the center in the decomposition $\Lie{u($3$)}=\Lie{u($1$)} \oplus \Lie{su($3$)}$ that determines this completely symmetric (bosonic) representation of \Lie{su($3$)}  to be characterized by the single non-negative integer $n$. 

Within the set of \Lie{su($3$)} generators \rf{su3gens} we find the subset that generates the rotation subalgebra $\Lie{so($3$)}\subset\Lie{su($3$)}$ of self-adjoint operators
\begin{equation}
\begin{aligned}
	{\cal L}_{x} &:= -\ii({\cal C}_y^z - {\cal C}_z^y)
	=\phantom{-\ii}{\textstyle\frac1{\surd2}}[(\eta_+{+}\eta_-)\xi_0 + \eta_0(\xi_+{+}\xi_-)],\\
	{\cal L}_{y} &:= -\ii({\cal C}_z^x - {\cal C}_x^z)
	=-\ii{\textstyle\frac1{\surd2}}[(\eta_+{-}\eta_-)\xi_0 - \eta_0(\xi_+{-}\xi_-)],\\
	{\cal L}_{z} &:= -\ii({\cal C}_x^y - {\cal C}_y^x)
	= \phantom{-\ii}\eta_+\xi_+-\eta_-\xi_-,
\end{aligned}\lab{so3ensu33}
\end{equation}
whose commutation relations are  
$[{\cal L}_{i},{\cal L}_{j}]=-\ii\,{\cal L}_{k}$ (with $i,\,j,\,k$ a cyclic permutation of $x,\,y,\,z$). The usual raising and lowering operators in \Lie{so($3$)} are
\begin{equation}
\begin{aligned}
	{\cal L}_+&:={\cal L}_x+\ii{\cal L}_y
			= \sqrt2(\eta_+\xi_0 - \eta_0\xi_-) ,\\
	{\cal L}_-&:={\cal L}_x-\ii{\cal L}_y
			= \sqrt2(\eta_-\xi_0-\eta_0\xi_+) , 
\end{aligned}\lab{lowrL}
\end{equation}
with $({\cal L}_{+})^\dagger = {\cal L}_{-}$, and commutators $[{\cal L}_{z},{\cal L}_{\pm}]=\pm{\cal L}_{\pm}$ and $[{\cal L}_+,{\cal L}_-]= -2{\cal L}_0\equiv -2{\cal L}_z$. The invariant Casimir operator is 
\begin{equation}
		{\cal L}^2={\cal L}_x^2+{\cal L}_y^2+{\cal L}_z^2
			= {\cal L}_\pm{\cal L}_\mp + {\cal L}_0({\cal L}_0\pm 1).
\end{equation}\lab{Casimirso3}

The three commuting generators 
\begin{equation}
	{\cal C}_x^x=\eta_x\xi_x,\quad 
	{\cal C}_y^y=\eta_y\xi_y,\quad 
	{\cal C}_z^z=\eta_z\xi_z, 
\end{equation}\lab{threeCs}
yield the three labels of the $\tau_{n}$ Cartesian states $\ket{n_{x}, n_{y}, n_{z}}$, while the invariant ${\cal C}$ in \rf{Ccenter}, ${\cal L}^2$ in \rf{Casimirso3}, and ${\cal L}_{z}$ in \rf{so3ensu33} determine the labels of the spherical basis $\roundket{n,\ell,m}$ with the ranges specified in \rf{varranges} and shown in Fig.\ \ref{fig:dos-cubos}. The $+$ — $-$ mode \emph{difference} $m:=n_{+} - n_{-}$, as in the $2D$ case, will have the role of angular momentum projection along the $z$--axis, \emph{provided} that the $\tau_{n}$ points in set can be fitted into \emph{complete} $\Lie{so($3$)}\supset\Lie{so($2$)}$ multiplets characterized by ranges $m|_{-\ell}^\ell$,i.e., those given by \rf{varranges}.

Figure \ref{fig:dos-cubos} also provides a geometric proof that shows that there are complete \Lie{so($3$)} multiplets in any one of the levels $n|_{0}^{N-1}$ in the lower `inverted pyramid' of finite-cube states. The lowest two rungs in the triangle of the figure contain $n + 1$ and $n$ points respectively; when projected, on the $m = \onehalf(n_{x} - n_{y})$ axis they sum $2n + 1$ points equidistant by unity, \emph{as if} they were a multiplet of highest angular momentum $\ell=n$. Then come the next higher two rungs of $n - 1$ and $n - 2$ points, that project on $2n - 3$ equidistant points, as if they belonged to a multiplet $\ell=n - 2$. In the triangle of Fig.\ \ref{fig:dos-cubos} the process continues with every two rungs yielding  $\ell=n - 4,\,n - 6$, etc., ending with $\ell=1$ if $n$ is odd or, when $n$ is even, the projection of the single apex point, $m=0$ of $\ell=0$. Of course, only the two extreme pairs of points $\pm\mu=\ell=n$ and $\pm(m - 1)$ are \emph{single} and belong to that highest $\ell$; all other $m$'s will be linear combinations of the $\ell$'s that we projected out geometrically in this figure.     

The symmetry importation on $0\le n\le N - 1=2j$ consists in using the coefficients that bind the  quantum harmonic oscillator Cartesian and spherical states; namely, the coefficients \rf{Elbracket} given explicitly by Chac\'on and de Llano \cite{Chacon-deLlano}. The slices in the upper pyramid of the cube of modes, for $4j=2(N - 1)\le n\le 3(N - 1)=6j$, yield reflected multiplets of `anti'-states, for which we may expect extra phases.

\begin{figure}[htbp]
\centerline{\includegraphics[width=0.9\columnwidth]{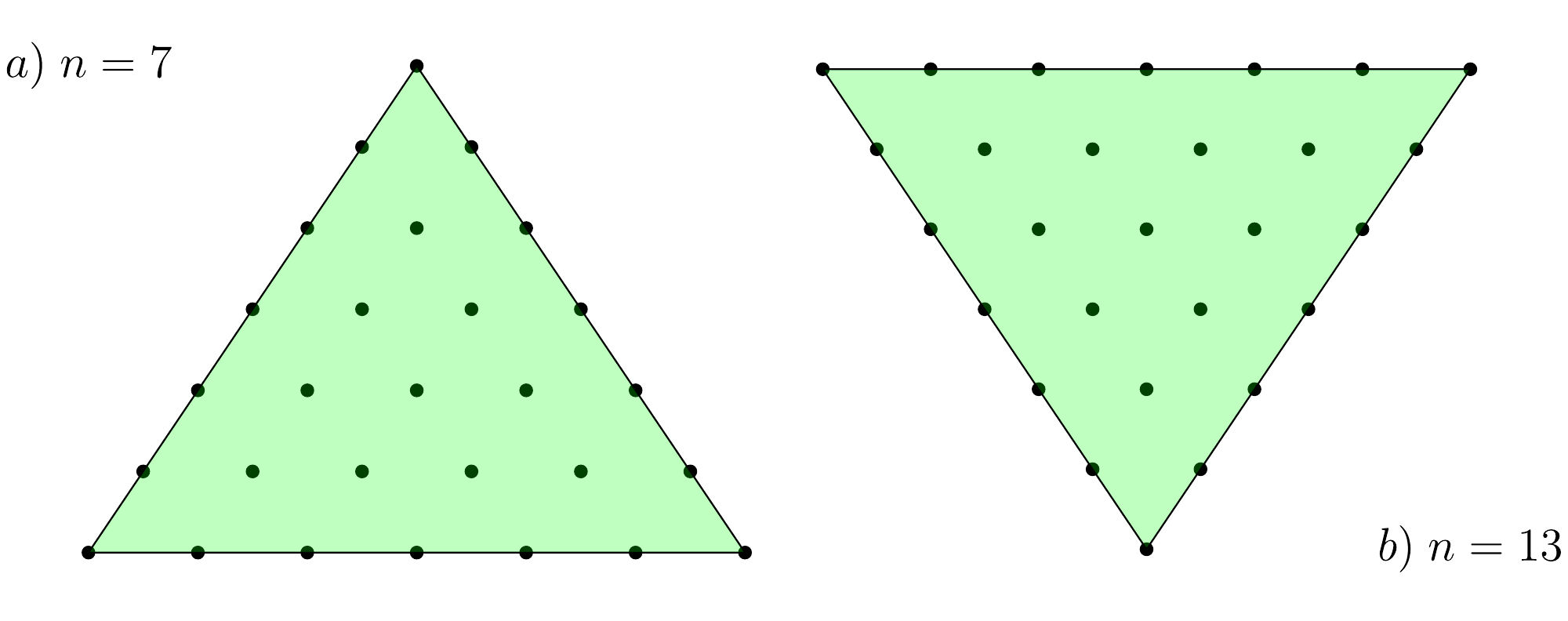}}
\caption[]{Mode states of the $3D$ finite oscillator, $\ket{n_x,n_y,n_z}_3$, $n_i|_0^{N-1}$ ($N=2j{+}1$), for total mode $n=\sum_{i} n_{i}$, form triangles of side $n+1$ and irregular hexagons that depends on the level $n$ reached. This slices the cube in Fig. \ref{fig:dos-cubos}. a) At the lower levels $n\in\{0,1,\ldots,2j\}$ of that figure  and, b) rotating by 180$^\circ$ those within  the highest range, $n\in\{4j,\,4j{+}1,\ldots,6j\}$. When we \emph{import} the $\Lie{su($3$)}\supset\Lie{so($3$)}$ symmetry contained in \rf{su3gens}--\rf{so3ensu33}, we  project out \emph{complete} multiplets $\roundket{\ell,m}$ of ${\cal L}^2$ and ${\cal L}_z$, where $\ell$ and $m$ have the ranges \rf{varranges}.}
\label{fig:triangulo}
\end{figure}

\begin{figure}[htbp]
\centerline{\includegraphics[width=0.9\columnwidth]{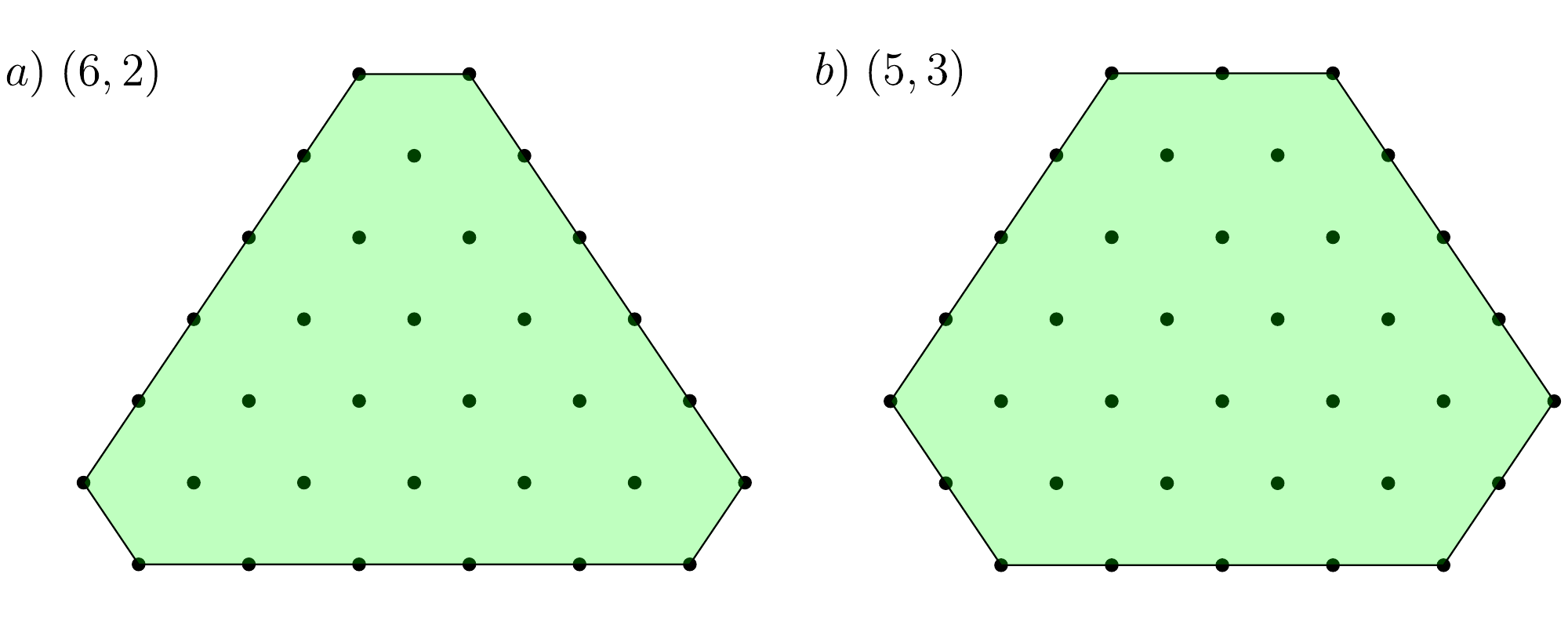}}
\caption[]{Hexagonal slices of the mode state space $\ket{n_x,n_y,n_z}_3$ of the $3D$ finite oscillator: a cube of $N=7$ points on the side (corresponding to $j=3$) and having $7^3=343$ points. a) For total mode numbers $n=8,\,9$ we show the hexagon $(6,2)$ and b) $(5,3)$. For $n=11,\,12$ the hexagons $(3,5)$ and $(2,6)$ are 180$^\circ$-rotated versions of these. Notice that $n=10$ has $(4,4)$, because we are dealing with an odd number of points, there exist a middle symmetrical hexagonal plane}
\label{fig:multipletes-hexagonales}
\end{figure}

The $3D$ case is more complicated than the $2D$ one, because we also have to consider the \emph{intermediate} range of total modes $N = 2j + 1\le n \le 4j - 1 = 2N - 3$. There, the $n$-slices of the mode cube are hexagons of generally unequal sides. We indicate by $(B,T)$ a hexagon with $B$ points on the base and $T$ points on the top. In particular, the $N$ triangles in the bottom pyramid ($n\in\{0,1,\ldots,N - 1\}$) are $(n + 1, 1)$ in Fig.\ \ref{fig:dos-cubos}, and the $N$ inverted triangles in the top pyramid $(1, 6j - n + 1)$. The $N - 2$ intermediate hexagons are thus $(N - 1, 2)$, $(N - 2, 3)$, $\ldots$, $(3, N - 2)$, $(2, N - 1)$. We have thus to ask whether the sets of points $(n_{x}, n_{y}, n_{z})$ in this intermediate region can be linearly combined into complete angular momentum multiplets, as those in the pyramid of Fig.\ \ref{fig:dos-cubos} ---or not.  In Fig.\ \ref{fig:multipletes-hexagonales} we show these intermediate slices for $N=7$ (a cube of $7^{3}=343$ voxels) that form the hexagons $(6,2)$ and $(5,3)$ for $n=6$ and $7$; the next two hexagons, $(3,5)$ and $(2,6)$ for $n=8$ and $9$, can be seen rotating the figure by 180$^\circ$. We note that we can indeed project out \emph{complete}  multiplets $\roundket{\ell,m}$ of eigenstates where $m=\onehalf(n_{x} - n_{y})|_{-\ell}^\ell$, with linear combinations of  $\ell\in\{5,\,4,\,2\}$ in the first hexagon and $\ell\in\{5,\,4,\,3\}$ in the second. As long as $N$ is even, we can scale up the figures to see that beyond the $(N,1)$ triangle at total mode $n=N - 1$, in the first intermediate hexagon $(N - 1,\,2)$ for $n=N$, the projected points will accommodate themselves into angular momentum multiplets $\ell\in\{N - 1,\,N - 2,N - 4,\ldots, 2\}$. For mode $n=N + 1$, the hexagon $(N - 2,\,3)$ will contain \Lie{so($3$)} multiplets $\ell\in\{N - 1,\,N - 2,\,N - 3,\,N - 5,\ldots,3\}$, and generally for $n=N + M$, the hexagon $(N - M + 1, M)$ will contain $\ell\in\{N - 1,\,N - 2,\ldots,\,N - M\} \cap\{N - M - 2,\,N - M - 4,\ldots,M\}$.

\section{Analysis of rotations}\label{sec:four}
\begin{figure}[htbp]
    \centering
    \includegraphics[width = \columnwidth]{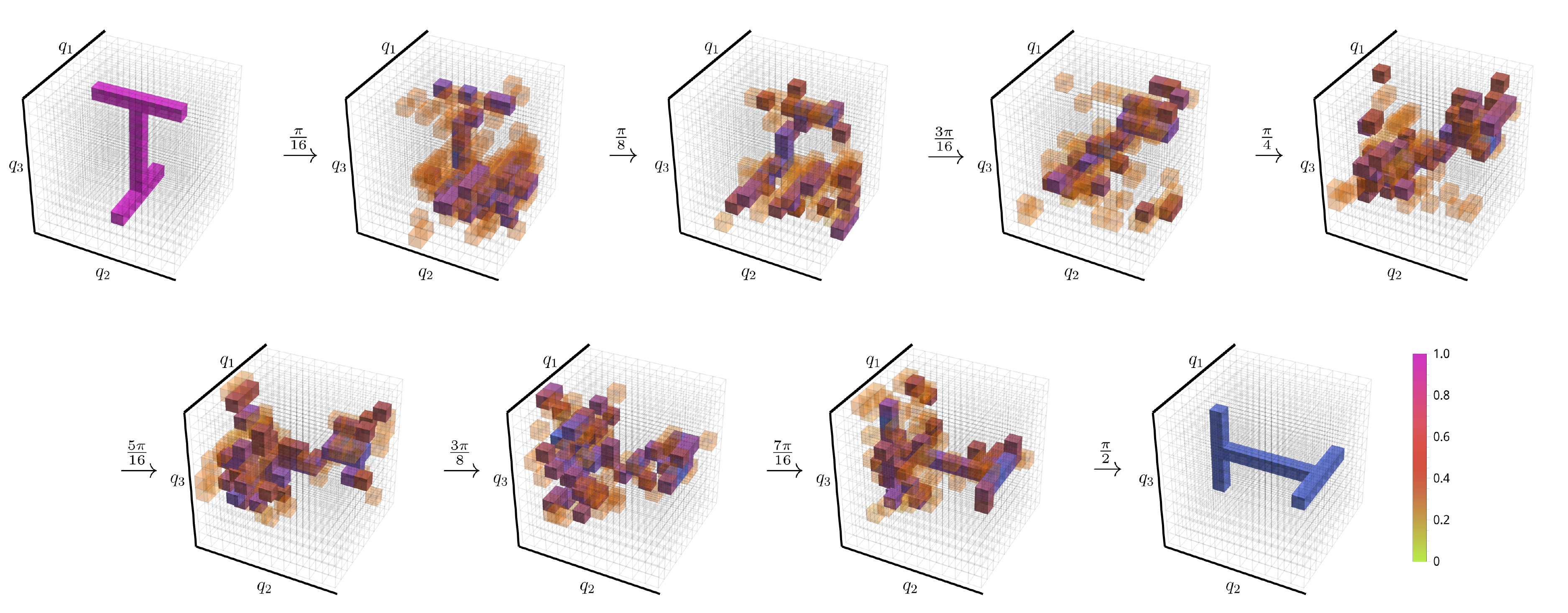}
    \caption{Rotation of a \emph{double-T}, that is, a T on top and bottom facing opposite directions. The three-dimensional space has dimension of $9^3$ voxels. The rotation is done around the $q_{1}$ and $q_{2}$ axes simultaneously, from $0$ to $\pi/2$, using the coefficients \rf{Elbracket} and the rotation kernel \rf{SO3trasf}. Here we see a spiral-like path from a vertical to a horizontal position of the volumetric figure.}
    \label{fig:doubleT_grid}
\end{figure}

In Figure \ref{fig:doubleT_grid} we show the composition of rotations for a ``volumetric pixelated image''. The example presented is based on the analytic expression for the coefficients \rf{Elbracket} and the rotation kernel \rf{SO3trasf}. There we have a \emph{double-T}, that is, a figure resembling the letter ``T'' facing opposing directions. From left to right, we concatenate rotations around axes $q_{1}$ and $q_{2}$, simultaneously; we choose steps $\Delta = \pi/16$, from $0$ to $\pi/2$. In this configuration, the Euler angles $\alpha=\beta$ and $\gamma$, define a spiral-like path around $q_{3}$, transitioning from vertical to horizontal position, and rotating the \emph{legs} of the double-T. At $\alpha=\beta=\pi/4$ the rod points toward the direction vector $n = (1,1,1)$ of the pixelated cube. 

The voxels show false color since the amplitude of every one of them has been scaled to be within the range $[0,1]$, using methods previously reported \cite{Urzua_2022}, where the \emph{principal} voxels that form the body of the rod are presented in a bluish tone, and the voxels with orangish tone are pixels appearing due the ``discrete Gibbs-\emph{like} phenomena" of the sharp edges in the transformation. We want to remark that, since the transformation is unitary when the rotation is performed, every one of the voxels acquires some amplitude value; but the voxels ranging below a threshold far from the body of the rod are treated as residues, and thus they are not shown in the figure, although they are taken into account when the transformation is sequentially performed.

\section{Conclusions}\label{sec:five}

The use of the Cartesian-to-spherical transformation coefficients of Bargmann and Moshinsky has been applied in optics before, by Pei and Liu \cite{Pei-Liu} for the purpose of efficiently approximating $3D$ Cartesian data through solid spherical harmonics under various cutoffs; subsequent work has used them to estimate rotation angles between Cartesian data \cite{Pei-Liu2}.

Here we extend the three-dimensional rotations previously reported at \cite{KUW}, were due to a limitation of the group of rotations importation, there was no way to freely rotate a set of voxels under concatenation. In this work, this limitation is overcome by the coupling between Cartesian and spherical basis, whose conjecture is proved to be accurate, since the composition of the transformation is truly unitary. Moreover, we don't need to perform some previous treatment to the original data, like smoothing or filtering, to overpass the limitations discussed above. In the sense of unitarity, as the set of previous works always stated, this method is extremely slow since every voxel depends on all the others, thus we are facing a computational cost of $~\mathcal{O}(N^6)$. 

Finally, the Fourier group is but a subgroup of the most general group \Lie{U($N^D$)} of unitary transformations among the $N^D$ pixel elements of images $f_{m_1,\ldots,m_D}\in{\cal C}^{N^D}$. This is the `aberration group' described in Ref.\ \cite{Valladolid} for 1D finite signals. At present we see no compelling application for this group beyond the two-dimensional case. Yet it would further the understanding of the structure of all transformations that conserve information in finite discrete systems, from the same viewpoint where linear and nonlinear canonical transformations conserve the structure of Hamiltonian geometric optics.

\section*{Funding}

Consejo Nacional de Ciencia y Tecnolog\'ia Postdoctoral Grant 2021.

\section*{Acknowledgements}

A.R. Urzúa thanks ICF UNAM for the resources provided during the postdoctoral stay. A.R. also wants to make the present work a homage and a memorial to Prof. Bernardo Wolf, who sadly passed away during the preparation of the manuscript.

\appendix
\section{Counting modes in the cube}
Here we give the detailed number of modes in each plane of the space $\{n_{x}+n_{y}+n_{z} = n$. This show the completeness of the method developed in this work since at the end, we must have the same number of points $N^{3}$ when we count the coplanar modes.

For the sake of clarity, we present three tables, the first and second are worked examples for $N=5, 7$, and the third is a generalization for $N = \#\mathrm{odd}$. The first column is the counting mode $n\in\{0, 3N-3\}$, and the second column shows the number of modes in the planes depicted as the base $B$ and tops $T$ in the pair $(B, T)$, the third column shows the number of angular momentum $\{l\}$ in the plane, and in the final column, we give the total number of modes that are coplanar at each level $n$.

\newtheorem{conj}{Conjecture}
\begin{conj}
The number of angular momenta over the whole cube is
\begin{equation*}
    \{\ell\}^{N}\;\{\ell_{max}-1\}^{N}\;\cdots\;\{1\}^{N},\;\; \mathrm{with}\;\; \ell_{max} = N-1,\;\; \mathrm{and}\;\; \{\ell\}^{N} :=N(2\ell+1).
\end{equation*}
\end{conj}

From this last, we can calculate the total number of points in the cube in terms of the angular momenta as
\begin{equation*}
\begin{aligned}
    \sum\limits_{l=0}^{N-1}\{\ell\}^{N} &= N(2N+1) + N(2N-3) + \cdots + 3N + N\\
        &= N\sum\limits_{M=1}^{N}(2N-2M+1)\\
        &= 2N^{3}-2N\sum\limits_{M=1}^{N}(M+N^{2})\\
        &= 2N^{3}-2N\left(\frac{1}{2}N(N+1)\right) + N^{2}\\
        &= N^{3}.
\end{aligned}
\end{equation*}

\begin{table}[htbp]
\centering
$
\begin{array}{ |c|c|c|c| }
 \hline
 \multicolumn{4}{|c|}{} \\
 \multicolumn{4}{|c|}{N = 5} \\
 \multicolumn{4}{|c|}{} \\
 \hline
 \hphantom{5em}n\hphantom{5em} & \hphantom{5em}(B, T)\hphantom{5em} & \hphantom{5em}\{\ell\}\hphantom{5em} & \hphantom{5em}\sum(2\ell+1)\hphantom{5em} \\
 \hline
 0 & (1, 1) & \{0\} & 1 \\
 1 & (2, 1) & \{1\} & 2 + 1 = 3 \\
 2 & (3, 1) & \{2, 0\} & 3 + 3 = 6  \\
 3 & (4, 1) & \{3, 1\} & 4 + 6 = 10 \\
 4 & (5, 1) & \{4, 2, 0\} & 5 + 10 = 15  \\
 \hline
 5 & (4, 2) & \{5, 3\} & 15 + 3 = 18 \\
 6 & (3, 3) & \{6, 4, 2\} & 18 + 1 = 19 \\
 7 & (2, 4) & \{5, 3\} & 15 + 3 = 18 \\
 \hline
 8 & (1, 5) & \{4, 2, 0\} & 15 \\
 9 & (1, 4) & \{3, 1\} & 10 \\
 10 & (1, 3) & \{2, 0\} & 6  \\
 11 & (1, 2) & \{1\} & 3 \\
 12 & (1, 1) & \{0\} & 1 \\
 \hline
 & & & \\
 & & \textrm{Total}: &  125 = 5^3 = N^3 \\ 
 \hline
\end{array}
$
\caption{For $N = 5$, we have a total of $N^3 = 125$ points in the cube. For every coplanar level $n$, we have a subset of points that has symmetrical reflection around $n = 6$, where $(B, T) = (3, 3)$. The sum of the total angular momentum ${l}$ leads to the total number of points $N^{3}$.}
\end{table}

\begin{table}[htbp]
\centering
$
\begin{array}{ |c|c|c|c| } 
 \hline
 \multicolumn{4}{|c|}{} \\
 \multicolumn{4}{|c|}{N = 7} \\
 \multicolumn{4}{|c|}{} \\
 \hline
 \hphantom{5em}n\hphantom{5em} & \hphantom{5em}(B, T)\hphantom{5em} & \hphantom{5em}\{\ell\}\hphantom{5em} & \hphantom{5em}\sum(2\ell+1)\hphantom{5em} \\
 \hline
 0 & (1, 1) & \{0\} & 1 \\
 1 & (2, 1) & \{1\} & 2 + 1 = 3 \\
 2 & (3, 1) & \{2, 0\} & 3 + 3 = 6  \\
 3 & (4, 1) & \{3, 1\} & 4 + 6 = 10 \\
 4 & (5, 1) & \{4, 2, 0\} & 5 + 10 = 15  \\
 5 & (6, 1) & \{5, 3, 1\} & 6 + 15 = 21 \\
 6 & (7, 1) & \{6, 4, 2, 0\} & 7 + 21 = 28 \\
 \hline
 7 & (2, 6) & \{5, 3, 1\} & 5 + 28 = 33 \\
 8 & (3, 5) & \{8, 6, 4, 2\} & 3 + 33 = 36 \\
 9 & (4, 4) & \{9, 7, 5, 3\} & 1 + 36 = 37 \\
 10 & (3, 5) & \{8, 6, 4, 2\} & 3 + 33 = 36  \\
 11 & (2, 6) & \{5, 3, 1\} & 5 + 28 = 33 \\
 \hline
 12 & (1, 7) & \{6, 4, 2, 0\} & 28 \\
 13 & (1, 6) & \{5, 3, 1\} & 21 \\
 14 & (1, 5) & \{4, 2, 0\} & 15  \\
 15 & (1, 4) & \{3, 1\} & 10 \\
 16 & (1, 3) & \{2, 0\} & 6 \\
 17 & (1, 2) & \{1\} & 3 \\
 18 & (1, 1) & \{0\} & 1 \\
 \hline
 & & & \\
 & & \textrm{Total}: &  343 = 7^3 = N^3 \\ 
 \hline
\end{array}
$
\caption{For $N = 57$, we have a total of $N^3 = 343$ points in the cube. For every coplanar level $n$, we have a subset of points that has symmetrical reflection around $n = 9$, where $(B, T) = (4, 4)$. The sum of the total angular momentum ${l}$ leads to the total number of points $N^{3}$.}
\end{table}

\begin{table}[htbp]
\centering
$
\begin{array}{ |c|c|c|c| } 
 \hline
 \multicolumn{4}{|c|}{} \\
 \multicolumn{4}{|c|}{N = \mathrm{odd}} \\
 \multicolumn{4}{|c|}{} \\
 \hline
 \hphantom{5em}n\hphantom{5em} & \hphantom{5em}(B, T)\hphantom{5em} & \hphantom{5em}\{\ell\}\hphantom{5em} & \hphantom{5em}\sum(2\ell+1)\hphantom{5em} \\
 \hline
 0 & (1, 1) & \{0\} & 1 \\
 1 & (2, 1) & \{1\} & 2 + 1 = 3 \\
 \cdots & \cdots & \cdots & \cdots  \\
 n & (n+1, 1) & \{n, n-2, ..., 0\;\textrm{or}\;1\} & \tau_{n} = \frac{1}{2}(n+1)(n+2) \\
 \cdots & \cdots & \cdots & \cdots  \\
 N-2 & (N-1, 1) & \{N-2, N-4, 0\} & \tau_{N-2} \\
 N-1 & (N, 1) & \{N-1, N-3, 1\} & \tau_{N-1} \\
 \hline
 N & (N-1, 2) & \{N-1, N-2, ..., 2\} & \frac{1}{2}(N^{2}+3N+2) \\
 N+1 & (N-2, 3) & \{N-1, N-2, ..., 3\} & \frac{1}{2}(N^{2}+5N-6) \\
 \cdots & \cdots & \cdots & \cdots \\
 \frac{3}{2}(N-1) & (\frac{N+1}{2}, \frac{N+1}{2}) & \{N-1, N-2, ..., \frac{N-1}{2}\} & \frac{1}{4}(3N^{2} + 1)  \\
 \cdots & \cdots & \cdots & \cdots \\
 2N-4 & (3, N-2) & \{N-1, N-2, ..., 3\} & \frac{1}{2}(N^{2}+5N-6) \\
 2N-3 & (2, N-1) & \{N-1, N-2, ..., 2\} & \frac{1}{2}(N^{2}+3N+2) \\
 \hline
 2N-2 & (1, 7) & \{N-1, N-3, 1\} & \tau_{N-1} \\
 2N-1 & (1, 6) & \{N-2, N-4, 0\} & \tau_{N-2} \\
 \cdots & \cdots & \cdots & \cdots  \\
 3N+3-n' & (1, n'+1) & \{n', n-2, 1\;\textrm{or}\;0\} & \tau_{n'} \\
 \cdots & \cdots & \cdots & \cdots \\
 3N-4 & (1, 2) & \{1\} & 3 \\
 3N-3 & (1, 1) & \{0\} & 1 \\
 \hline
 & & & \\
 & & \textrm{Total}: &  N^{3} \\ 
 \hline
\end{array}
$
\caption{Generalization for the counting modes when we deal with a cube of $N^{3}$ odd number of points. We see that, in the same fashion as the $2D$ case for integer representation size $j$, we have a central mode at $\frac{3}{2}(N-1)$. For semi-integer $j$, there exist two middle points at $\frac{3}{2}(N-2)$ and $\frac{3}{2}(N-1)$.}
\end{table}

\newpage
\bibliographystyle{unsrt}
\bibliography{bib.bib}

\end{document}